\documentclass{aa}
\usepackage{graphicx}

\begin{document}
\title{A substellar companion around the \\ intermediate-mass giant star HD~11977 
\thanks {Based on observations collected at the 1.52-m ESO telescope from
  October~1999 to February~2002 and at the 2.2-m MPG/ESO telescope from
  December~2003 to November~2004 at the La Silla Observatory, Chile.}
}

\author{J. Setiawan\inst{1}, J. Rodmann\inst{1}, L. da Silva\inst{2}, 
A.~P. Hatzes\inst{3}, L. Pasquini\inst{4}, \\
O. von der L\"uhe\inst{5}, J.~R. de Medeiros\inst{6}, M.~P. D\"ollinger\inst{4}
and L. Girardi\inst{7}
\\
} 
\offprints{J. Setiawan, \texttt{setiawan@mpia.de}}
\institute{Max-Planck-Institut f\"ur Astronomie, K\"onigstuhl 17, 69117, Heidelberg, Germany
\and
Observat\'orio Nacional, R. Gal. Jos\'e Cristino, 77, 20921-400 S\~ao Crist\'ov\~ao, Rio de Janeiro, Brazil
\and
Th\"uringer Landessternwarte, Sternwarte 5, 07778, Tautenburg, Germany
\and
European Southern Observatory, Karl-Schwarzschild-Str. 2, 85748, Garching bei M\"unchen, Germany
\and
Kiepenheuer-Institut f\"ur Sonnenphysik, Sch\"oneck-Str. 6, 79104, Freiburg, Germany
\and
Departamento de F\'\i sica, Universidade Federal do Rio Grande do Norte, 59072-970 Natal, Brazil
\and
Osservatorio Astronomico di Trieste, via Tiepolo 11, 34131 Trieste, Italy
}

\date{Received 2 February 2005/Accepted 25 May 2005}

\abstract{
We report the discovery of a substellar companion to the intermediate-mass
star \object{HD~11977} (G5\,III). Radial velocities of this star have been 
monitored for five years with FEROS at the 1.52-m~ESO and later at the 
2.2-m~MPG/ESO telescope in La Silla, Chile. Based on the collected data we 
calculated an orbital solution with a period of \mbox{$P=711\thinspace\rm{days}$}, 
a semi-amplitude of \mbox{$K_{1}=105\thinspace\mathrm{m\,s}^{-1}$}, 
and an eccentricity of \mbox{$e=0.4$}. The period of the 
radial-velocity variation is longer than that of the estimated stellar rotation, 
rendering it unlikely that rotational modulation is the source of the 
variation in the radial velocity. This hypothesis is supported by the 
absence of a correlation between stellar activity indicators and 
radial-velocity variation. 
By determining a primary stellar mass of \mbox{$M_{\star}=1.91\thinspace\rm{M_{\sun}}$}, 
the best-fit minimum mass of the companion and semi-major axis of the orbit are

\mbox{$m_{2}\sin{i}=6.54\thinspace\rm{M_{Jup}}$} and 
\mbox{$a_{2}=1.93\thinspace \rm{AU}$}, respectively. An upper limit for the mass of
the companion of \mbox{$m_{2} \la 65.5\thinspace\mathrm{M_{Jup}}$} has
been calculated from {\sc Hipparcos} astrometric measurements. 
Although the possibility of a brown-dwarf companion cannot be excluded, 
HD~11977~B is one of the few planet candidates 
detected around an intermediate-mass star. 
The progenitor main-sequence star of \object{HD~11977} is probably an A-type star. 
This discovery gives an indirect evidence for planetary companions around 
early type main-sequence stars.

}

\authorrunning{J. Setiawan et al.}
\titlerunning{A substellar companion around the intermediate-mass giant star HD 11977}
\maketitle
%
%________________________________________________________________
\section{Introduction}
Among the substellar companions discovered around other stars, 
there are only a few planetary and brown-dwarf companions to
intermediate-mass stars \mbox{($M_{\star}\approx$~1.5--7$~\mathrm{M}_{\sun}$)}. 
In fact, the presence of planetary or brown-dwarf companions 
to stars in this mass regime is not well explored
in current radial-velocity planet search programmes. 
Ongoing radial-velocity surveys, e.g. CORALIE~\&~HARPS survey 
and California~\&~Carnegie Planet Search, concentrate on solar-like stars. 
Only few planetary companions have been detected around 
evolved stars, like G and K~giants (Hatzes~et~al.~\cite{hat05} 
and references herein).

Intermediate-mass main-sequence stars have spectral classes from A to early F
and higher effective temperatures than solar-type stars. Therefore,
their spectra show fewer absorption lines than later, solar-like 
spectral types (late F, G and K). Due to faster stellar rotation 
\mbox{($v\sin{i}>30\,\mathrm{km\,s}^{-1}$)} their line profiles are also broader.
These facts make it difficult to determine accurately the stellar radial
velocity~(RV).

Cool evolved stars, on the other hand, are suitable targets for precise RV measurements.
As a star moves towards the red giant branch (RGB), its rotational velocity 
decreases and the stellar atmosphere cools significantly. 
Red giants, however, are chromospherically active (Pasquini~et~al.~\cite{pas00}). 
The variation of the chromospheric activity of G-K~giants leaves 
an imprint in the spectral line profiles. Rotational modulation 
due to starspots can also produce RV variations, thereby mimicking 
the gravitational influence of low-mass companions. This effect can be 
investigated through the variations in the spectral line profile asymmetry 
and by monitoring the chromospheric activity indicators, 
like \mbox{Ca\,II~K and H} emission lines, see e.g., Choi~et~al~(\cite{cho95}).
Non-radial oscillations may also induce RV variations as shown 
by the asteroseismology of several giant stars (Hatzes~\&~Cochran~\cite{hat94}; 
Buzasi~et~al.~\cite{buz00}; Frandsen~et~al.~\cite{fra02}).

In this letter we report the discovery of a substellar companion
around \object{HD~11977}. We have excluded rotational modulation as the source of 
the RV variations, as described in the following sections. 

\section{About the star HD~11977}

\object{HD~11977} is a cool evolved star of spectral type G5\,III ({\sc Hipparcos}, ESA~1997). 
In the Hertzsprung-Russell diagram \mbox{($M_{V}$ vs. T$_{\mathrm{eff}}$)} it is located in the 
``clump region'' of the RGB. After ascending the RGB, the star moved down to the 
clump region where it is undergoing helium-core and hydrogen-shell burning.
Clump giants typically spend $10^{8}$ years in this evolutionary stage, 
that, for initial masses higher than \mbox{1.7~M$_{\sun}$}, represents 
more than 5\% of their main-sequence lifetime (Girardi~et~al.~\cite{gir00}). 

\begin{table}[h]
\caption{Stellar parameters of \object{HD~11977}} 
\begin{tabular}{lll}
\hline\hline\\[-3mm]
Spectral type	    		& G5III \\
$m_{V}$   			& 4.68  	      & mag   \\
$M_{V}$     		    	& $0.57  \pm 0.07$    & mag   \\
$B-V$		                & $0.931 \pm 0.036$   & mag   \\		 
Parallax		        & $15.04 \pm 0.47$    & mas   \\
Distance		        & $66.5  \pm 2.1$     & pc    \\
Effective temperature $T_{\mathrm{eff}}$ & $4970  \pm 70$    	  & K\\
Stellar radius R$_{\star}$ 	& $10.09  \pm 0.32$     & R$_{\sun}$ \\
Angular diameter	        & $1.49  \pm 0.02$      & mas    \\
Metallicity [Fe/H]		& $-0.21 \pm 0.1$       & dex     \\
Surface gravity $\log\big(g/\mathrm{cm\,s}^{-2}\big)$ & $ 2.90  \pm 0.2$	  &    \\
Stellar mass $M_{\star}$        & $1.91  \pm 0.21$  	& M$_{\sun}$ \\ 
$\log\,\mathrm{(Age/year)}$	& $8.7-9.3$        &    \\
$v_{\mathrm{rot}}\sin{i}$	& $2.4\pm1.0$       & km\,s$^{-1}$ \\
Micro turbulence	        & $1.40-1.60$      & km\,s$^{-1}$ \\
$P_{\mathrm{rot}}/\sin{i}$  	& $230-270$        & days   \\[0.5mm]
\hline
\hline
\end{tabular}
\end{table}

\begin{figure}[ht]
\begin{center}
\includegraphics[width=8.5cm,height=8.0cm]{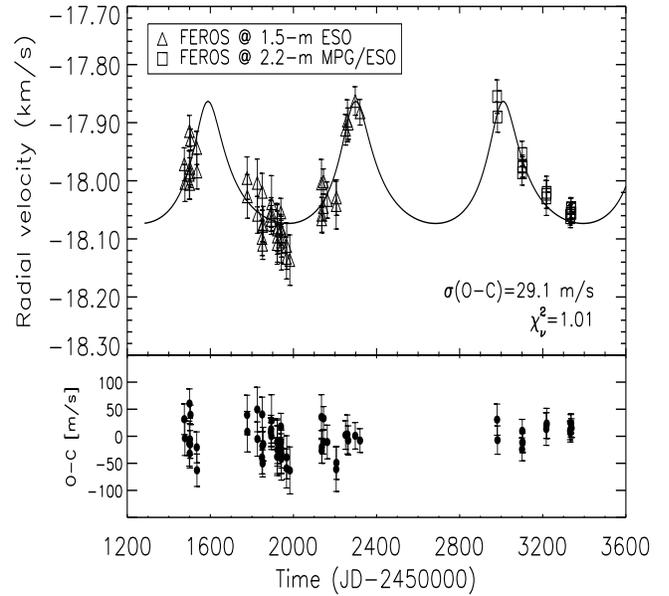}
\caption{Radial velocity measurements of HD~11977, taken with FEROS from 
October~1999 to November~2004. The measurements can be fitted with an orbital 
solution with a period of 711~days and an eccentricity of 0.4. 
The lower panel shows the residual velocity. The rms scatter of the residuals
is partly due to intrinsic variability on short time scales, as commonly found
for giant stars.}
\end{center}
\end{figure}

The basic stellar parameters of \object{HD~11977} are listed in Table~1. 
The spectral type, photometric data and parallax were taken from the 
{\sc Hipparcos} catalogues.
Adelman~(\cite{ade01}) reported a photometric variability of 0.5~millimag.  
Thus, \object{HD~11977} is one of the least variable stars known from the {\sc Hipparcos} 
photometric survey.
 
The solar-metallicity evolutionary tracks by Girardi~et~al.~(\cite{gir00}) 
show that the mass of \object{HD~11977} is, indicatively, 
between 2.5--3.0~M$_{\sun}$. 
In a forthcoming paper (da Silva~et~al., in preparation) the $T_{\mathrm{eff}}$, 
$R_{\star}$, [Fe/H] and $\log{g}$ are measured from the spectra of \object{HD~11977}. 
Their values are given in Table~1. Then, using as a constraint the observed $M_V$, $B-V$, 
[Fe/H] values and their uncertainties, and the Girardi~et~al. theoretical isochrones, 
we derived the mass probability distribution function for \object{HD~11977} in a 
more rigorous way.
The subsolar metallicity of this star ($\mathrm{[Fe/H]}=-0.21\pm0.1$) 
lead us to derive a mass of $1.91\pm0.21$~M$_{\sun}$, 
and to exclude masses above 2.5~M$_{\sun}$.
A star of such mass and metallicity is not expected to have suffered
significant mass loss during its ascent of the RGB. 
Therefore, the progenitor of \object{HD~11977} was probably an A-type main sequence star.

The projected rotational velocity $v\sin{i}$ was computed using the 
cross-correlation method (e.g., Benz~\&~Mayor~\cite{ben84}; Setiawan~et~al.~\cite{set04}). 
From the projected rotational velocity and the 
stellar radius we calculated the upper limit of the rotation period as 
\mbox{$P_{\mathrm{rot}}/\sin{i}$~=~230--270~days}. 
Such a procedure of $P_{\mathrm{rot}}$ determination is 
very sensitive on the $v\sin{i}$ precision. 
Nevertheless, the $v\sin{i}$ value we have obtained for 
\object{HD~11977} is in excellent agreement with the value expected for 
a solar-type G5III star with $B-V\sim 0.90$, $v\sin{i}= 2.1\,
\mathrm{km\,s}^{-1}$ (de Medeiros~et~al.~\cite{dem96}), pointing 
for an excellent precision on our $v\sin{i}$ determination.

\section{Observations and analysis}
We have monitored the radial velocity of \object{HD~11977} from October 1999 
to February 2002 with the echelle spectrograph FEROS \mbox{($R=\lambda/\Delta\lambda=48\,000$)}
at the \mbox{1.52-m~ESO} telescope. We continued the monitoring at the \mbox{2.2-m~MPG/ESO} 
telescope, to which FEROS had been moved, from December 2003 to November 2004.

FEROS is equipped with two fibres and can be operated in ``object-sky'' and 
``object-calibration'' mode (Kaufer~\&~Pasquini~\cite{kau98}). 
We used the simultaneous calibration technique (Baranne~et~al.~\cite{bar96}) and 
the cross-correlation method to determine the radial velocity of \object{HD~11977}. 
Details of the data reduction and RV computation 
procedures have been described in Setiawan~et~al.~(\cite{set03}).
The radial velocity measurements of \object{HD~11977} are 
shown in Fig.\,1.

\section{Evidence for a substellar companion}

We observed a long-period variation in the radial velocity of \object{HD~11977}.
A Lomb-Scargle periodogram (Scargle \cite{sca82}) of the data yields significant 
power at a frequency of \mbox{0.0014 cycles\,day$^{-1}$} (Fig.\,2) and a false alarm 
probability (FAP) of $\sim$10$^{-9}$ using expressions given in Scargle~(\cite{sca82}).
This low FAP value was confirmed using the bootstrap randomisation technique 
(K\"urster~et~al.~\cite{kue97}) and 200\,000 random shuffles of the data
which yielded a \mbox{FAP $< 5 \times 10^{-6}$}, i.\,e. no power in the
random data greater than the data power.

\begin{figure}[h]
\begin{center}
\includegraphics[width=7.5cm, height=6cm]{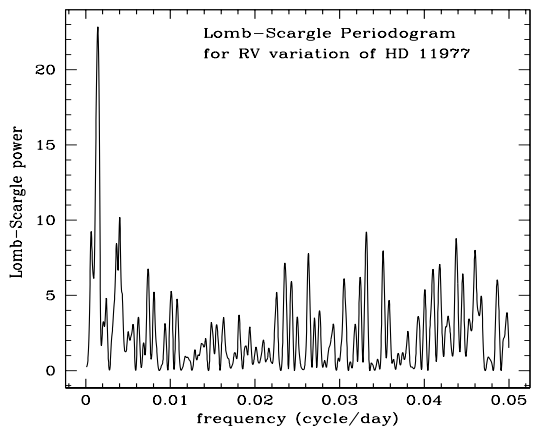}
\caption{The Lomb-Scargle periodogram of the radial velocity measurements. 
The periodogram shows a significant power at a frequency of \mbox{0.0014 cycles\,day$^{-1}$}.}
\end{center}
\end{figure}

\begin{figure}[h]
\begin{center}
\includegraphics[width=7.5cm, height=6.0cm]{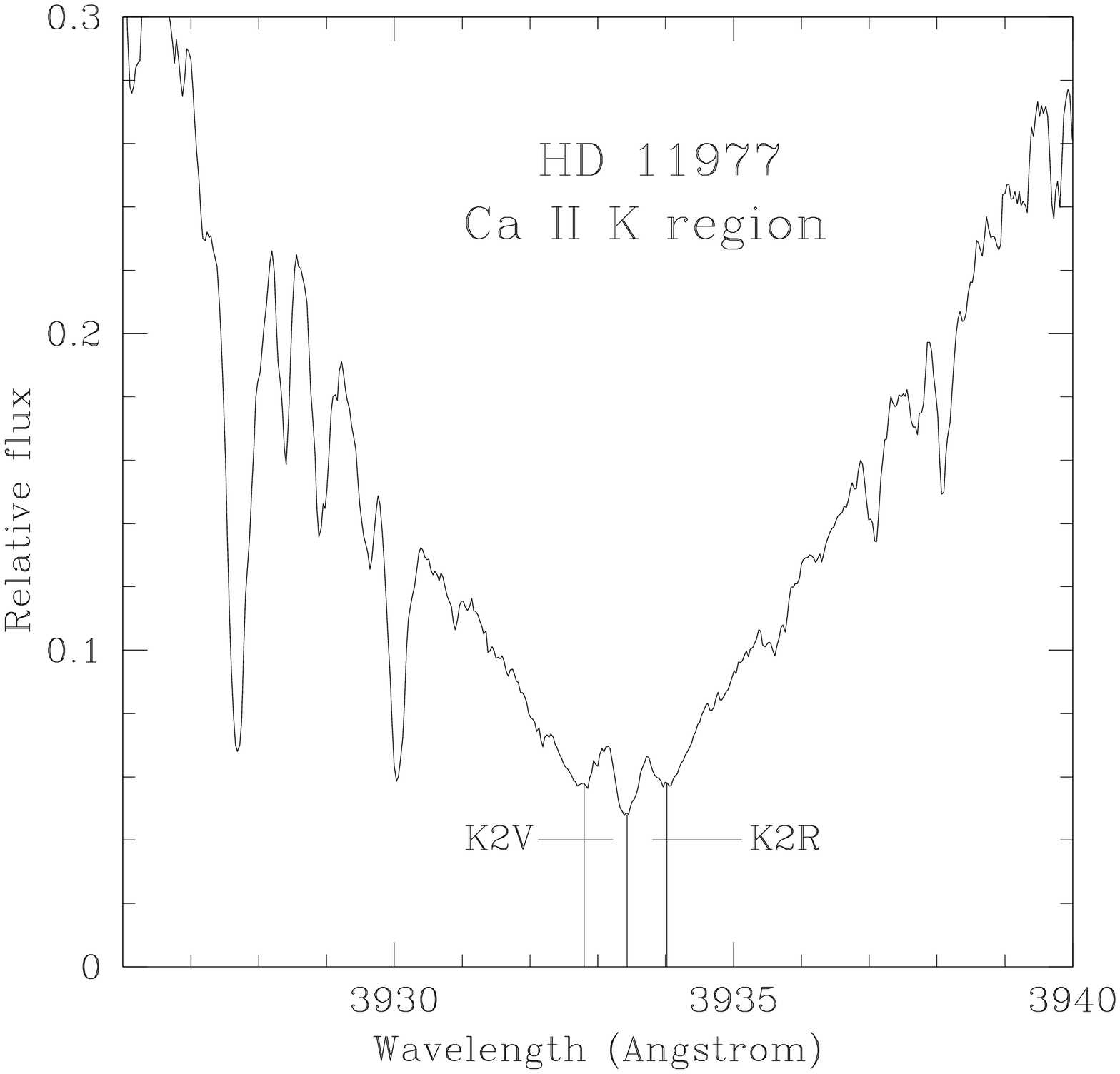}
\includegraphics[width=7.5cm, height=6.0cm]{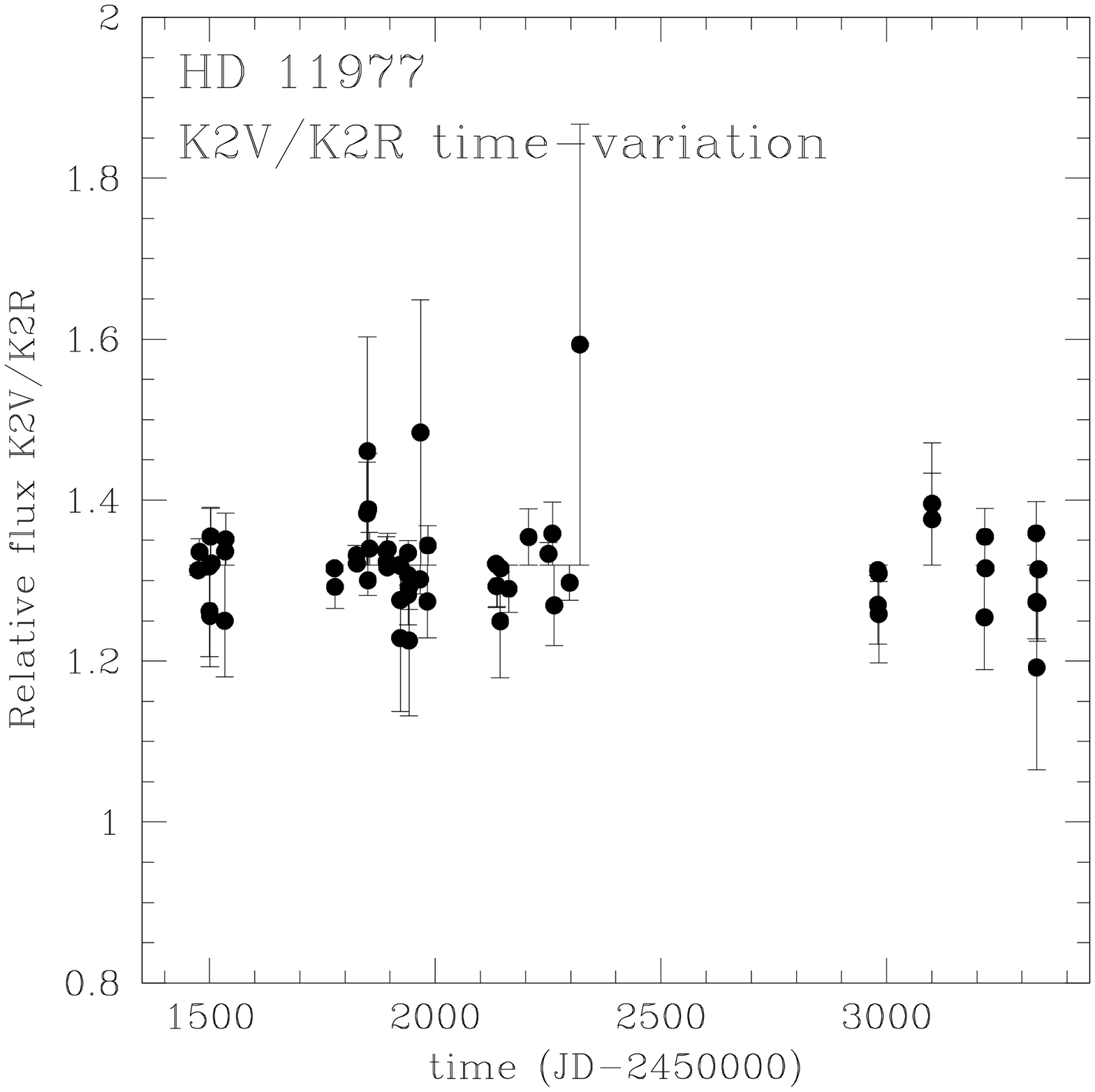}
\caption{The upper panel shows the region around the Ca~II~K emission lines. 
The relative flux K2V/K2R (lower panel) shows no long-period variation.}
\end{center}
\end{figure}

It is unlikely that the RV variation of \object{HD~11977} is due to rotational 
modulation. The upper limit of the rotation period ($P_{\mathrm{rot}}/\sin{i}<300$~days) 
is obviously much lower than the observed period in the RV variability. 
The analysis of the Ca~II~K emission lines (Fig.\,3) did not show any 
significant stellar activity (e.g., due to surface inhomogeneities). 
We measured the relative flux K2V/K2R. 
This method has been applied on several red giants to detect rotational 
modulation (Setiawan~et~al.~2005).
In addition, we measured the bisector velocity span (e.g. Queloz~et~al.~\cite{que01}) and 
found no dependence between the asymmetry of the spectral line profile and the measured 
radial velocities (Fig.\,4). 

\begin{figure}[h]
\centering
\includegraphics[width=7.5cm, height=6.0cm]{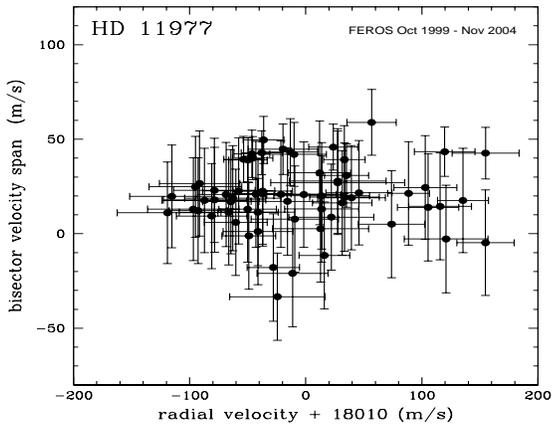}
\caption{Bisector velocity span (BVS) of \object{HD 11977} calculated 
from the spectra taken between October 1999 and November 2004. 
The diagram shows no correlation between BVS and the measured radial velocities.}
\end{figure}

Assuming a semi-amplitude of 105~$\mathrm{m\,s}^{-1}$ and 
a rotational velocity of 2.4~$\mathrm{km\,s}^{-1}$, 
a simple starspot model results in a spot-filling factor of 
4\% of the stellar hemisphere (Hatzes~\cite{hat02}). 
For a starspot close to the equator and with $\Delta T$~=~1200~K, 
this translates to a photometric variation of $\sim$0.1~mag 
(Launhardt~\&~Setiawan, in preparation), 
which is of a factor of $\sim$200 larger than the value given in 
Adelman~(\cite{ade01}). 
However, we cannot exclude the possibility that the {\sc Hipparcos} 
photometric measurements were done at a time when \object{HD~11977} was in
the ``quiet'' phase of its activity cycle.

The astrometric variation of \object{HD~11977} is less than 
$1\,$mas ({\sc Hipparcos}). This gives a lower limit for \mbox{$\sin{i}\ga 0.1$}. 
If the $\sin{i}$ were 0.1, the true rotation velocity would be about \mbox{24~$\mathrm{km\,s}^{-1}$}. 
A K~giant rotating this fast would be very active. 
In this case, strong Ca~II~K emission lines should be expected.  
We did not observe this feature in our spectroscopic measurements. 

We therefore exclude rotational modulation due to surface inhomogeneities 
as the source of the RV variation, and infer the presence of a substellar 
companion (planet or brown dwarf) orbiting \object{HD~11977}. 
The orbital solution to these data yielded a period of $P=$~711~days, 
a semi-amplitude of $K_{1}=$~105~$\mathrm{m\,s}^{-1}$ and 
an eccentricity of $e=$~0.4. The residual of \mbox{29.1 $\mathrm{m\,s}^{-1}$} is 
comparable with the expected long-term accuracy of FEROS.
The computed orbital parameters yield a minimum mass of the companion  
\mbox{$m_{2}\sin{i}=$~6.54~M$_{\mathrm{Jup}}$} (Table~2).

\begin{table}[h]
\caption{Orbital parameters for \object{HD~11977~B}} 

\vspace{-0.5cm}
$$
\begin{array}{lclcl}
\hline\hline\\[-3mm]
P		                && 711\pm8                    && \mathrm{days}  \\
T_{0}\mathrm{\;(JD-2450000)}   	&& 1420                       && \mathrm{days}  \\
e      		                && 0.4\pm0.07                 &&                \\
V		                && -18.010                    && \mathrm{km\,s}^{-1} \\
\omega_{1}	                && 351.5\pm9.5                && \mathrm{deg}   \\
K_{1}		                && 0.105\pm0.008              && \mathrm{km\,s}^{-1} \\
\sigma(O-C)	                && 29.1                       && \mathrm{m\,s}^{-1}  \\
f(m)		                && (6.55\pm1.63)\times10^{-8} && \mathrm{M}_{\sun}            \\
a_{1}\sin{i} 	                && 9.41\times10^{5}&& \mathrm{km}       \\[0.5mm]
\hline\\[-3mm]
\mathrm{with} \, m_{1}=M_{\star}=\,$1.91$\,\mathrm{M}_{\sun}        \\
m_{2}\sin{i} 	                && $6.54$                      && \mathrm{M}_{\mathrm{Jup}} 	\\
a_{2} 		                && $1.93    $                 && \mathrm{AU}       \\[0.5mm]
\hline
\hline
\end{array}
$$
\end{table}

The {\sc Hipparcos} astrometry data yields an upper limit for the companion mass 
of less than 66~M$_{\mathrm{Jup}}$, i.\,e.\ below or near the hydrogen-burning limit. 
The expectation value of \mbox{$\sin{i}=\pi/4$} for randomly oriented orbital 
planes can be used to derive a value of $m_{2}$. We derived a value of 
\mbox{$m_{2} \approx 8.3~\mathrm{M_{Jup}}$}.  Since this value is below the deuterium-burning 
limit ($\sim$13~M$_{\mathrm{Jup}}$), \object{HD~11977~B} 
is one of very few planetary companions discovered around intermediate-mass giant stars. 

\section{Discussion}

To date there are only two substellar companions detected around intermediate-mass stars, 
namely HR~7329 (Lowrance~et~al.~\cite{low00}) and HD~13189 (Hatzes~et~al.~2005). 
HR~7329~A is a main-sequence star of spectral type A0 of mass $M_{\star}=2.9$~M$_{\sun}$. 
HR~7329~B is a brown-dwarf companion with $m_{2}\approx$~30--50~M$_{\mathrm{Jup}}$ 
in an orbit of $\sim$200~AU (Guenther~et~al.~\cite{gue01}). 
HD~13189~A is a K~giant/supergiant of mass $M_{\star}=$~2--7~M$_{\sun}$. 
The distance to this star is only poorly known, thus its stellar mass 
cannot be determined accurately. 
This leads to large uncertainty for the mass of the substellar companion. 
The companion's minimum mass was estimated between 8--20~M$_{\mathrm{Jup}}$. 
Due to this uncertainty HD~13189~B can be either a planet or brown dwarf. 

%Because of the poor statistics of substellar companion orbiting intermediate-mass stars, 
%it is still not clear whether the low number of detections is due 
%to instrumental limitations or differences in the planet formation process 
%around intermediate-mass stars. Close substellar (planet/brown-dwarf) companions to main-sequence 
%intermediate-mass stars or even massive stars are very difficult to detect with the 
%current techniques. 
%On the one hand, only young and therefore self-luminous substellar companions 
%with large, resolvable separations to the parent star can be detected directly 
%with current imaging instruments. On the other hand, ``normal'' 
%early-type main-sequence stars are ``too hot'' and have only a few, 
%broad absorption lines which are not suitable for accurate RV determination. 
%Only when these stars have evolved to slow rotating giants with cooler atmosphere, 
%can their RV be measured precisely. Due to this limitation, 
%current precise RV surveys focus on solar-type stars (Fig.\,5).

\begin{figure}[ht]
\flushleft\includegraphics[width=9.0cm, height=7cm]{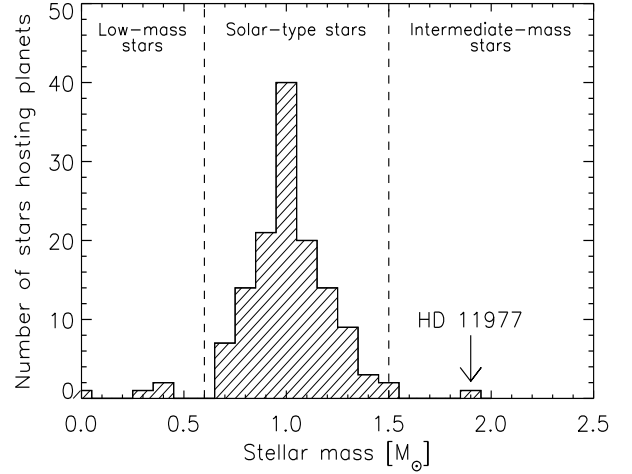}
\caption{Mass distribution of stars with known planetary companions 
(as of April 2005, including \object{HD~11977}). We exclude HD~13189 
from this histogram due to the large uncertainty of the stellar mass.} 
\end{figure}

\object{HD~11977} is the first intermediate-mass 
star with accurate mass around which a planetary companion has been discovered
(Fig.\,5). 
Assuming negligible mass loss during the RGB~phase, the main-sequence 
progenitor of \object{HD~11977} is probably an A star with an 
stellar mass of approximately $2$~M$_{\sun}$ (Schmidt-Kaler~\cite{sch82}).
\object{HD~11977} has subsolar metallicity (\mbox{[Fe/H]=$-$0.21}). 
Low metal abundance has been also found in other giants hosting planets 
(Sato~et~al.~\cite{sat03}; da Silva~et~al.~in preparation). 
This point will be investigated in more detail in our next 
study of stellar abundances. 

Our discovery gives a crucial support to search for extrasolar planets 
around early type main-sequence stars and stars in late evolutionary stages. 
Detections of planetary companions to stars outside the solar-mass regime 
can provide valuable constraints for theories of planet formation. 

\begin{acknowledgements}
We thank John Pritchard and Fernando Selman for their works on FEROS.
For the long-period assistance during the observations with the 1.52-m ESO 
we thank Rolando Vega and Arturo Torrejon. 
We also thank Francisco Labrana, Mauro Stefanon, Karla Aubel and Manuel Pizzaro for the
supports during the observations at the 2.2-m MPG/ESO.
\end{acknowledgements}


\begin{thebibliography}{}

\bibitem[2001]{ade01} Adelman, S.J. 2001, A\&A, 367, 297

\bibitem[1996]{bar96} Baranne, A., Queloz, D., Mayor, M., et al. 1996, A\&AS, 119, 373.

\bibitem[1984]{ben84} Benz, W., \& Mayor, M. 1984, A\&A, 138, 183.

\bibitem[2000]{buz00} Buzasi, D., Catanzarite, J., Laher, R. et al. 2000, \apj, 532, 133

\bibitem[1995]{cho95} Choi, H-J., Soon, W., Donahue, R.A. et al. 1995, \pasp, 107, 744

%\bibitem[2005]{sil05} da Silva et al., {\it in preparation}

\bibitem[1996]{dem96} de Medeiros, J.R., da Rocha, C., \& Mayor, M. 1996, A\&A, 314, 499

%\bibitem[1997]{esa97} ESA  1997, The Hipparcos and and Tycho Catalogues

\bibitem[2002]{fra02} Frandsen, S., Carrier, F., Aerts, C. et al. 2002, A\&A, 394, 5

\bibitem[2000]{gir00} Girardi, L., Bressan, A., Bertelli, G. et al. 2000, A\&AS, 141, 371.

\bibitem[2001]{gue01} Guenther, E.~W., Neuh{\"a}user, R., Hu{\'e}lamo, N. et al. 2001, \aap, 365, 514

%\bibitem[1993]{hat93} Hatzes, A.~P. \& Cochran, W.~D. 1993, \apj, 1993, 413, 339

\bibitem[1994]{hat94} Hatzes, A.~P. \& Cochran, W.~D. 1994, \apj, 422, 366

\bibitem[2002]{hat02} Hatzes, A.~P., 2002, AN, 323, 392

\bibitem[2005]{hat05} Hatzes, A.~P., Guenther, E.~W., Endl, M. et al. 2005, A\&A, in press

\bibitem[1998]{kau98} Kaufer A. \& Pasquini, L., 1998, SPIE, 3355, 844

\bibitem[1997]{kue97} K\"urster, M.,  Schmitt, J.H.M.M., Cutispoto, G. et al. 1997, A\&A, 320, 831

%\bibitem[2005]{lau05} Launhardt, R. \& Setiawan, J., {\it in preparation}

\bibitem[2000]{low00} Lowrance, P.~J., Schneider, G., Kirkpatrick, J.~D. et al., 2000, \apj, 541, 390

%\bibitem[1998]{mar98} Marcy, G., \& Butler, P., 1998, ARAA, 36, 57

\bibitem[2000]{pas00} Pasquini, L., de Medeiros, J.~R. \& Girardi, L., 2000, A\&A, 361, 1011

%\bibitem[2001]{pas01} Pasinetti Fracassini, L.~E., Pastori, L., Covino, S. et al. 2001, A\&A, 367, 521 

\bibitem[2001]{que01} Queloz, D., Henry W., Sivan, J.P., et al. 2001, A\&A, 379, 279

%\bibitem[2005]{ric05} Richichi, A., Percheron, I., Khristoforova, M. 2005,
  A\&A, 431, 773

%\bibitem[2004]{sch04} Sch\"onberner, D., {\it private communication}

\bibitem[2003]{sat03} Sato, B., Hiroyashu, A., Eiji, K. et al. 2003, \apj, 597, L157

\bibitem[1982]{sca82} Scargle, J.D., 1982, ApJ, 263, 835

\bibitem[1982]{sch82} Schmidt-Kaler, Th., 1982, in Landolt-B\"ornstein Numerical Data and Functional 
Relationships and Technology, Vol.~2, Schaifers/Vogt, Springer-Verlag Berlin

\bibitem[2003]{set03} Setiawan, J., Pasquini, L., da Silva, L., et al. 2003, A\&A, 397, 1151.

\bibitem[2004]{set04} Setiawan, J., Pasquini, L., da Silva, L. et al. 2004, A\&A, 421, 241.

\bibitem[2005]{set05} Setiawan, J., von der L\"uhe, O., da Silva, L. et al. 2005, in proceeding of the 
Cool Stars, Stellar Systems and the Sun 13, held in Hamburg, Germany 5--7 July 2004, in press


%% Old references, mainly about brown dwarfs

%\bibitem[2003]{clo03} Close, L.~M., Siegler, N., Freed, M. et al. 2003, \apj, 587, 407
%\bibitem[2001]{fis01} Fischer, D.~A., Marcy, G.~W., Butler, R.~P. et
%al. 2001, \apj, 551, 1107
%\bibitem[2002]{fis02} Fischer, D.~A., Marcy, G.~W., Butler, R.~P. et al. 2002,
  %\pasp, 114, 529
%\bibitem[2002]{fis03} Fischer, D.~A. et al. 2003
%\bibitem[1996]{flo96} Flower, P.J. 1996, ApJ, 469, 355.
%\bibitem[2003]{fre03} Freed, M., Close, L.~M., \& Siegler, N. 2003, \apj, 584, 453
%\bibitem[2001]{giz01} Gizis, J.~E., Kirkpatrick, J.~D., Wilson, J.~C. et al. 2001, \aj, 121, 2185
%\bibitem[2001]{gue01} Guenther, E.~W., Neuh{\"a}user, R., Hu{\'e}lamo, N. et
%al. 2001, \aap, 365, 514
%\bibitem[2002]{jon02} Jones, H.~R.~A., Paul Butler, R., Tinney, C.~G. et al. 2002, \mnras, 333, 871
%\bibitem[1999]{low99} Lowrance, P.~J., McCarthy, C., Becklin, E.~E. et al. 1999, \apjl, 512, L69
%\bibitem[2000]{low00} Lowrance, P.~J., Schneider, G., Kirkpatrick, J.~D. et al., 2000, \apj, 541, 390
%\bibitem[1996]{per96} Perryman et al. 1996
%\bibitem[1998]{que98} Queloz, D., Allain, S., Mermilliod, J.C., et al. 1998,
%A\&A, 335, 183.
%\bibitem[1998]{reb98} Rebolo, R., Zapatero Osorio, M.~R., Madruga, S. et al. 1998, Science, 282, 1309
%\bibitem[2002]{san02} Santos, N.C., PhD Thesis, Observatoire de Geneve, 2002


\end{thebibliography}
\end{document}